\documentclass{article}
\usepackage{spconf,amsmath,graphicx,hyperref}
\usepackage{url}
\usepackage{booktabs}
\usepackage{multicol}
\usepackage{multirow}
\usepackage{listings}
\usepackage{amsmath}
\usepackage{amssymb}
\usepackage{enumitem}
\usepackage{cite}
\usepackage[capitalise]{cleveref}
\usepackage{color}

\title{MuseTok: Symbolic Music Tokenization for Generation \\ and Semantic Understanding}
\name{
\begin{tabular}{c}
    Jingyue Huang \qquad 
    Zachary Novack \qquad 
    Phillip Long \qquad 
    Yupeng Hou \\
    Ke Chen \qquad 
    Taylor Berg-Kirkpatrick \qquad 
    Julian McAuley
\end{tabular}
}
\address{
    University of California San Diego, USA
}

\begin{document}
\ninept
\maketitle
\begin{abstract}
Discrete representation learning has shown promising results across various domains, including generation and understanding in image, speech and language. Inspired by these advances, we propose MuseTok, a tokenization method for symbolic music, and investigate its effectiveness in both music generation and understanding tasks. MuseTok employs the residual vector quantized-variational autoencoder (RQ-VAE) on bar-wise music segments within a Transformer-based encoder-decoder framework, producing music codes that achieve high-fidelity music reconstruction and accurate understanding of music theory. For comprehensive evaluation, we apply MuseTok to music generation and semantic understanding tasks, including melody extraction, chord recognition, and emotion recognition. Models incorporating MuseTok outperform previous representation learning baselines in semantic understanding while maintaining comparable performance in content generation. Furthermore, qualitative analyses on MuseTok codes, using ground-truth categories and synthetic datasets, reveal that MuseTok effectively captures underlying musical concepts from large music collections.
\end{abstract}
\section{Introduction}\label{sec:introduction}
Discrete representation learning aims to train models to represent data within a finite set of discrete codes~\cite{VQ-VAE}. It has proven effective across diverse generative tasks, including image generation~\cite{VQ-VAE, VQ-VAE-2, RQ-VAE}, neural speech codec~\cite{SoundStream}, generative retrieval for recommender systems~\cite{TIGER} and signal-level music generation~\cite{Jukebox, MusicGen}.
In music information retrieval (MIR), discrete representations have also been applied to genre classification~\cite{JukeMIR} and melody transcription~\cite{sheetsage}. 
Such methods span a wide range of compression bottlenecks, from lightly compressed discrete codes for improved modeling~\cite{MusicGen}, to highly compressed codes capturing deep semantic information~\cite{agostinelli2023musiclm}.
However, work in discrete representation learning currently lags in the \emph{symbolic} music domain. While some limited previous research has explored the use of discrete representations for classification tasks~\cite{MIDI-BERT, MusicBERT} or controllable generation~\cite{wu2021musemorphose, Poly-Dis, figaro}, such work only focused on the specific application rather than on general representations for diverse tasks, with limited attention to \emph{how} one should learn semantic embeddings, nor to the \emph{quality} of such representations.

% == architecture ==
Thus, we introduce \textbf{MuseTok}, the first tokenization method for general symbolic music representations that can support multiple applications,  including symbolic music generation and semantic music understanding in multiple perspectives.
% a new tokenization method for symbolic music data based on discrete representation learning, supporting hierarchical symbolic music generation and semantic music understanding in multiple perspectives.
%
We leverage an encoder-decoder architecture with residual quantization~\cite{RQ-VAE} to learn bar-wise music residual codes through reconstruction, on top of music sequences derived by REMI+~\cite{figaro}. 
Analysis of code usage and similarities demonstrates its effectiveness in capturing music theoretical concepts, such as textures and musical intervals.

Regarding the applications, for music generation, we employ a Transformer decoder to predict MuseTok codes, then pass codes to another Transformer decoder to generate REMI+ events.
For semantic music understanding, three classification tasks are considered to assess the note-level, bar-level and song-level music semantics embedded in the codes.
We adopt public-domain symbolic music data for model training, including a large-scale dataset PDMX~\cite{PDMX, xu2024generating} and several small datasets~\cite{pop909, cp, emopia, piano-midi, hymnal, ragtime} spanning diverse genres, with a main focus on piano pieces to explore discrete representation learning in a single-instrument setting. 
Our contributions are three-fold: 
\begin{itemize}[leftmargin=*,itemsep=0pt,topsep=0pt]
    \item We propose MuseTok, the first discrete representation learning framework of symbolic music for general purpose, applicable to both generation and understanding tasks. 
    \item MuseTok achieves comparable performance on symbolic music generation and superior performance on two of three classification tasks to previous baselines, demonstrating its effectiveness on content generation and semantic understanding. 
    \item We provide analysis on how MuseTok learns underlying musical concepts, such as key, interval, time signature, and texture.
\end{itemize}

We present generation samples and more details about datasets and experiments on the website\footnote{\href{https://musetok.github.io/}{https://musetok.github.io/}}. Code implementation and checkpoints are open sourced\footnote{\href{https://github.com/Yuer867/MuseTok}{https://github.com/Yuer867/MuseTok}}.

\begin{figure*}[t]
 \centering
 \includegraphics[width=0.99\textwidth]{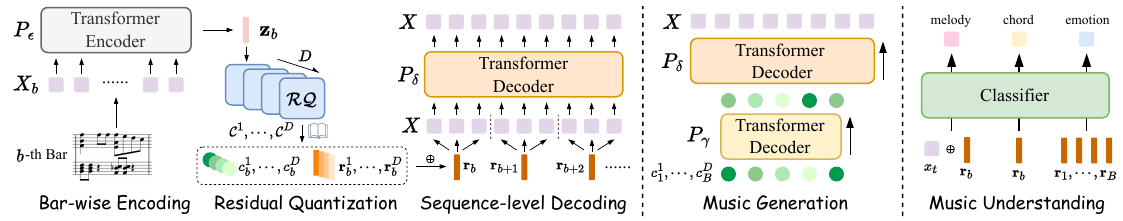}
 \vspace{-0.3cm}
 \caption{Overview of MuseTok (left) and its downstream generation (middle) and understanding (right) tasks.}
 \vspace{-0.2cm}
 \label{fig:overview}
\end{figure*}

\vspace{-0.2cm}
\section{Related Work}\label{sec:related_work}

\textbf{Music Representation Learning.}
The success of representation learning methods~\cite{BERT, CLIP, CLAP, VQ-VAE, VQ-VAE-2, RQ-VAE, SoundStream, MusicGen, TIGER, Jukebox} has inspired exploration in the symbolic music domain.
For music understanding, BERT-like models have been applied for classification tasks~\cite{MIDI-BERT, MusicBERT}, while contrastive methods integrate music with language~\cite{CLaMP}.
For generation, VAE-based models disentangle latent variables to encode attributes like chord and texture, enabling controllable generation~\cite{PianoTreeVAE, Poly-Dis} and style transfer~\cite{wu2021musemorphose, figaro}.
Unlike prior work targeting specific tasks, this paper focuses on general symbolic music representations, exploring their broad potential in music downstream tasks and quality.
\\

\vspace{-0.2cm}
\noindent\textbf{Symbolic Music Encoding.}
Various encoding formats have been proposed for symbolic music generation. 
MIDI-Message~\cite{musictransformer, PerformanceRNN} and REMI~\cite{remi} encode MIDI data as sequence of events like note, beat and time shift.
Later work introduced representations for compound attributes~\cite{cp}, %byte-pair words ~\cite{BPE-music, AnalyzingBPE, FromWordToMusic}, 
multiple instrument tracks~\cite{MMT, figaro, MMM}, expressive performance~\cite{StructuredMIDI, PerTok} and emotion control~\cite{emopia, EMO-Disentanger}.
This paper investigates discrete music representation learning on the bar-level segments, introducing a new tokenization for music generation and understanding.

\section{Methods}\label{sec:methods}

\subsection{Music Tokenization}
As illustrated in \cref{fig:overview} (left), we adopt the idea of RQ-VAE~\cite{RQ-VAE} to construct an encoder-decoder architecture with residual quantization (RQ) blocks to learn discrete representations of symbolic music.
\\

\vspace{-0.2cm}
\noindent\textbf{Encoder.} 
Symbolic music input is converted into a REMI+~\cite{figaro} sequence $X = \{X_1, \cdots, X_B\}$ over $B$ bars, where $X_b$ contains all REMI+ events within the $b$-th bar. A Transformer encoder $P_{\epsilon}$ processes each bar to produce latent embeddings $\mathbf{z}_1, \cdots, \mathbf{z}_B$.
\\

\vspace{-0.2cm}
\noindent\textbf{Residual Quantization.}
Residual quantization blocks $\mathcal{RQ}$ discretize each $\mathbf{z}_b$ into embeddings $\mathbf{r}$ and corresponding codes $c$ (indices) from codebooks $\mathcal{C}^1, \cdots, \mathcal{C}^D$:
\vspace{-0.05cm}
\begin{equation}
    (c_b^1, \mathbf{r}_b^1), \cdots, (c_b^D, \mathbf{r}_b^D) = \mathcal{RQ}(\mathbf{z}_b; \mathcal{C}^1, \cdots, \mathcal{C}^D)
\vspace{-0.05cm}
\end{equation}
where $D$ is the number of codebooks or quantization depth, ($c^d_b$,$\mathbf{r}^d_b$) is the retrieved code (index) and corresponding embedding in the $d$-th codebook $\mathcal{C}^d$ for $\mathbf{z}_b$. Each codebook $\mathcal{C}^d$ contains $K$ index-embedding pairs $\{(k^d, \mathbf{e}_k^d)\}_{k=1}^K$, where $k^d$ is the index and $\mathbf{e}_k^d$ is its corresponding embedding in $\mathcal{C}^d$, so $\mathbf{r}^d_b \in \{\mathbf{e}^d_k\}_{k=1}^K$, $c^d_b \in \{k^d\}_{k=1}^K$.

The first code $c_b^1$ of $\mathbf{z}_b$ is the index in the codebook $\mathcal{C}^1$ whose embedding $\mathbf{r}_b^1$ is nearest to $\mathbf{z}_b$.
Then $\mathcal{RQ}$ recursively computes codes $c_b^2 \cdots c_b^D$ so that their embeddings are nearest to the residuals:
\vspace{-0.05cm}
\begin{align}
    c_b^1 &= \operatorname{argmin}_k || \mathbf{z}_b - \mathbf{e}_k^1||, \\
    c_b^d &= \operatorname{argmin}_k || \mathbf{z}_b - \mathbf{e}_k^d - \Sigma_{i=1}^{d-1} \mathbf{r}^i_b ||, 2 \leq d \leq D.
    \vspace{-0.05cm}
\end{align}
To capture different granularities of music contents, the codes and embeddings in $D$ codebooks are not shared.
\\

\vspace{-0.2cm}
% RQ-VAE - decoder
\noindent\textbf{Decoder.} 
After obtaining all aggregated embeddings $\{\mathbf{r}_b$=$\Sigma_{d=1}^D \mathbf{r}_b^d | b=1,...,B\}$ from the $\mathcal{RQ}$ module, a Transformer decoder $P_{\delta}$ decodes these embeddings to predict the music sequence $X$ in an auto-regressive mode. The reconstruction objective is:
\vspace{-0.1cm}
\begin{equation}
    \mathcal{L}_{recon} = -\sum_{t=1}^T \operatorname{log}P_{\delta}(x_{t+1}|x_{\leq t}; \mathbf{r}_{\leq b}), b = \text{bar}(t)
\end{equation}
where $x_t$ denotes the $t$-th event in $X$, as $X =\{X_1, \cdots, X_B\} = \{x_1, \cdots, x_T\}$, and $\text{bar}(t)$ is the bar index where the $t$-th event lies.
\\

\vspace{-0.2cm}
\noindent\textbf{Codebook Utility.} 
To better utilize the codebook during training, we adopt SimVQ~\cite{SimVQ} and rotation trick~\cite{rotation_trick} to improve codebook utility and reconstruction quality, with the commitment objective as
\vspace{-0.1cm}
\begin{equation}
    \mathcal{L}_{commit} = \sum_{d=1}^D \sum_{b=1}^B \left|\left|\mathbf{z}_b -\operatorname{sg}\left[\sum_{d'=1}^d \mathbf{r}_b^{d'}W^d\right]\right|\right|_2^2
    \vspace{-0.1cm}
\end{equation}
where $\operatorname{sg}$ is the stop-gradient operator, $W^d$ is the linear transformation per codebook from SimVQ. 
The final training objective is the combination of two objectives: $\mathcal{L} = \mathcal{L}_{recon} + \mathcal{L}_{commit}$. Codebooks are updated by the exponential moving average~\cite{VQ-VAE}.

\subsection{Music Generation}

As shown in \cref{fig:overview} (middle), with the trained tokenization model, another Transformer decoder $P_{\gamma}$ is connected before the MuseTok decoder $P_{\delta}$ for two-stage music generation. 
Using code sequence ${c_1^1, \cdots, c_B^D}$ converted from REMI+ sequence by the frozen $P_{\epsilon}$ and $\mathcal{RQ}$, the generator $P_{\gamma}$ is trained to perform next-token prediction on codes with cross-entropy loss. 
During inference, $P_{\gamma}$ first generates MuseTok codes for high-level musical structure, then decoded by the frozen $P_{\delta}$ into REMI+ events for fine-grained music details.

\subsection{Music Understanding}

We employ three classification tasks to assess semantic understanding of MuseTok at the note, bar and song levels in \cref{fig:overview} (right).
\\

\vspace{-0.2cm}
\noindent\textbf{Melody Extraction.} The symbolic-domain melody extraction aims to identify melody notes from single-track polyphonic music~\cite{MIDI-BERT}.
A classifier is trained to assign each pitch event $x_t$
of $X$ to one of three classes, vocal melody, instrumental melody or accompaniment, using the code embedding $\mathbf{r}_b$ ($b = \text{bar}(t)$) as a condition to provide note-level semantic context, as in the MuseTok decoding process.
\\

\vspace{-0.2cm}
\noindent\textbf{Chord Recognition.}
We introduce a symbolic-domain chord recognition task that extracts chord progressions from single-track polyphonic music to evaluate the harmony information captured by MuseTok.
Given the code embedding $\mathbf{r}_b$ for a bar, a classifier predicts a chord label for each beat from a set of predefined categories.
\\

\vspace{-0.2cm}
\noindent\textbf{Emotion Recognition.}
Emotion recognition classifies a song into one of four categories defined by high/low positiveness and high/low activation~\cite{russell}, evaluating the song-level semantic capability of MuseTok.
For a given song, a classifier is trained to take the code embeddings $\mathbf{r}_1, \cdots, \mathbf{r}_B$ as input and predicts its emotion label.

\begin{table*}[t]
\small\centering
\resizebox{0.78\linewidth}{!}{
\begin{tabular}{l|ccc|ccc|c}
\toprule
                                & \multicolumn{3}{c|}{\textbf{PPL $\downarrow$}}      & \multicolumn{3}{c|}{\textbf{Acc (\%) $\uparrow$}}    & \multirow{2}{*}{\textbf{Util $\uparrow$}}\\
                                & mono.           & chora.           & poly.           & mono.           & chora.         & poly.         &               \\ 
\midrule
VAE                             & 1.123\footnotesize{$\pm$0.007} & 1.086\footnotesize{$\pm$0.007} & \textbf{1.159}\footnotesize{$\pm$0.021} & 96.70  & 83.59 & 82.21 & -            \\
\midrule
% Residual FSQ                    &                                &                                &                                &                 &               &               &               \\
MuseTok-Small                   & 1.093\footnotesize{$\pm$0.005} & 1.169\footnotesize{$\pm$0.021} & 1.498\footnotesize{$\pm$0.053} & 97.76  & 80.85 & 62.92 &  \textbf{99.58}\%      \\
\quad -- w/o. SimVQ+Rota.    & 1.151\footnotesize{$\pm$0.008} & 1.268\footnotesize{$\pm$0.027} & 1.667\footnotesize{$\pm$0.059} & 80.55  & 64.04 & 54.59 & 87.77\%      \\
\quad -- w/o. SimVQ             & 1.129\footnotesize{$\pm$0.006} & 1.242\footnotesize{$\pm$0.025} & 1.620\footnotesize{$\pm$0.057} & 85.87  & 66.19 & 56.15        &  90.50\%      \\
\quad -- w/. PDMX only          & 1.106\footnotesize{$\pm$0.005} & 1.174\footnotesize{$\pm$0.021} & -                              & 97.91  & 78.11 & -       &  99.29\%      \\ 
MuseTok-Large                   & \textbf{1.091}\footnotesize{$\pm$0.005} & \textbf{1.081}\footnotesize{$\pm$0.010} & 1.217\footnotesize{$\pm$0.030} & \textbf{99.58}  & \textbf{93.71} & \textbf{82.68} &  98.96\%      \\ 
\bottomrule
\end{tabular}
}
\vspace{-0.1cm}
\caption{Reconstruction performance across different texture groups and codebook utility for ablated versions of MuseTok.}
\vspace{-0.3cm}
\label{table:tokenization}
\end{table*}

% \begin{table}[t]
% \small\centering
% \resizebox{0.95\columnwidth}{!}{
% \begin{tabular}{lrrrc}
% \toprule
% \textbf{Dataset}        & \textbf{\# pieces}    & \textbf{\# bars}  & \textbf{\# events}    & \textbf{Genres}       \\
% \midrule
% PDMX-mono.\,\cite{PDMX, xu2024generating} &  163,366              &  27.36            &  19.33                &   mixed               \\
% PDMX-contra.\,\cite{PDMX, xu2024generating} &  25,597               &  35.24            &  44.77                &   mixed               \\
% POP909\,\cite{pop909}       &  886                  &  83.52            &  62.31                &   pop                 \\
% EMOPIA\,\cite{emopia}   &  1,071                &  17.79            &  46.73                &   pop                 \\
% Pop1k7\,\cite{cp}   &  1,747                &  104.59           &  39.39                &   pop                 \\
% Hymnal\,\cite{hymnal}&  1,606                &  18.47            &  47.05                &   folk                \\
% Multipianomide\,\cite{piano-midi}        &  457                  &  95.64            &  47.89                &   classical           \\
% Ragtime\,\cite{ragtime}               &  457                  &  139.16           &  49.32                &   jazz                \\
% \midrule
% \textbf{Summary}        &  195,187              &  29.21            &  24.30                &   -                   \\
% \bottomrule
% \end{tabular}
% }
% \vspace{-0.1cm}
% \caption{The datasets. The \#bars are averaged over a dataset, while \#events are averaged per bar.}
% \vspace{-0.3cm}
% \label{table:dataset}
% \end{table}

\section{Experiments}\label{sec:experiments}

\subsection{Datasets and Pre-processing}
% == datasets ==
We collect piano pieces from the large-scale PDMX~\cite{PDMX} and six small datasets covering diverse genres: POP909~\cite{pop909}, EMOPIA~\cite{emopia}, Pop1k7~\cite{cp}, Hymnal~\cite{hymnal}, Multipianomide~\cite{piano-midi} and Ragtime~\cite{ragtime}.
%\footnote{Dataset statistics are presented in the demo page.}. 
% == pre-processing ==
%
Pieces with time signature changes are segmented into clips with a consistent time signature and at least 8 bars.
Since most samples maintain constant tempo and velocity, these performance-related attributes are removed to focus on structural and harmonic aspects. 
To improve structural consistency across datasets, we align note onsets and durations to valid sheet music positions. % enforcing formal composition principles.
%
% == some statistics ==
After pre-processing and REMI+~\cite{figaro} encoding, the resulting 195,187 sequences (83.7\% monophonic, 13.1\% chorale and 3.2\% polyphonic) are randomly split into training, validation and test sets at an 8:1:1 ratio, yielding a vocabulary of 140 music events.

\subsection{Music Tokenization}\label{subsec:tokenization}

\textbf{Model Settings and Ablations.} 
The tokenization model uses 12-layer, 8-head Transformers with 512 hidden dim. for both encoder and decoder.
Training is performed on 16-bar sequences, augmented by random transposition within $\pm6$ semitones, with chorale and polyphonic pieces unsampled to balance texture groups.
Using a 128-dim VAE as the upper bound, we evaluate several residual quantization variants, including ablations on quantization techniques (Rotation trick, SimVQ), dataset usage (PDMX only) and model size: MuseTok-Small (quantization depth $D$=$8$, codebook size $K$=$1024$, 128-dim) and MuseTok-Large ($D$=$16$, $K$=$2048$).
All models are trained with Adam optimizer (learning rate $1\mathrm{e}{-4}$) with 200-step warm-up, converging in 45k steps on a single RTX A6000.
\\

\vspace{-0.2cm}
\noindent\textbf{Evaluation Metrics.}
We evaluate reconstruction quality and codebook utility using three metrics. 
Perplexity (PPL) measures how well a model predicts a sequence of music events, defined as inversely proportional to the log-probability of the test split. 
Accuracy (Acc) is computed as $1 - \mathcal{D}(X, X^\prime)/|X|$,
where $\mathcal{D}(X, X^\prime)$ refers to the edit distance between the reconstructed sequence $X^\prime$ and the original $X$.
Both metrics are evaluated on three texture groups: monophonic, chorale, and polyphonic, reflecting increasing musical complexity.
Codebook utility (Util) measures the fraction of codes used at least once when encoding the test set, averaged across quantization layers.
\\

\vspace{-0.2cm}
\noindent\textbf{Results.}
From \cref{table:tokenization}, most models exhibit highest perplexity and lowest reconstruction accuracy on polyphonic pieces, due to their complex textures and chord progressions.
Incorporating diverse datasets during training improves reconstruction on PDMX (mostly monophonic and chorale pieces) compared to training on PDMX alone, validating the effectiveness of balanced sampling and pre-processing in aligning dataset distributions.
Among MuseTok-Small variants, combining SimVQ with rotation tricks yields the highest codebook utility and best reconstruction quality.
MuseTok-Large further approaches or surpasses the VAE upper bound, particularly improving on polyphonic pieces over MuseTok-Small.
Based on these results, we select MuseTok-Large as the tokenization model for generation and understanding tasks. 

\subsection{Music Generation}\label{subsec:generation}

\textbf{Model Settings.}
The first-stage generator is a 12-layer, 16-head Transformer with 1024 hidden dim., totaling 152M parameters, trained with sequence length 256 using the same hyperparameters as tokenization, converging in $\sim$200k steps over 4 days.
The second stage adopts the trained tokenization decoder.
Datasets are augmented by offline key transposition ($\pm6$).
Unbalanced pieces are resampled during training.
During inference, nucleus sampling~\cite{sampling} is applied ($\tau$=$1.1$, $p$=$0.9$), followed by top-$k$ downsampling ($k$=$30$).
\\

\vspace{-0.2cm}
\noindent\textbf{Baselines.}
We compare two baselines of standard symbolic music encoding methods, a Transformer decoder trained on REMI+ sequence with the same datasets and model size as ours (REMI)~\cite{remi}, and Anticipatory Music Transformer (AMT)~\cite{Anticipatory} using provided \texttt{music-small-100k} checkpoint, on a music continuation task, which allows reliable assessments through relative comparison.
VAE-based models like MuseMorphose~\cite{wu2021musemorphose} and FIGARO~\cite{figaro} are not compared as they require reference music for generation.
% also adopt VAE or VQ-VAE for music feature extraction, they only work given reference music.
\\

\vspace{-0.2cm}
\noindent\textbf{Evaluation Metrics.}
We evaluate continuation results using two objective metrics and a subjective online listening test. 
To quantitatively evaluate the similarity to the primers, we adopt bar-wise chroma similarity $\operatorname{sim}_{\mathrm{chr}}$ and grooving similarity $\operatorname{sim}_{\mathrm{grv}}$~\cite{wu2021musemorphose}, measuring tonal closeness via cosine similarity of chroma vectors~\cite{Fujishima1999RealtimeCR} and rhythmic resemblance via grooving vectors~\cite{grooving}. 
Sequence-level similarity is computed by averaging the highest similarity scores between each generated bar and the primer bars.

In the listening test, based on 4-bar primers, participants rate generated continuations on a 5-point Likert scale for four aspects: Pitch (Pit.), Structure (Str.), Harmony (Har.), and Development (Dev.).
We collected 24 responses from participants spanning a wide spectrum of musical expertise, each evaluating 8 groups of random samples, yielding 192 ratings per model per aspect.
\\

\vspace{-0.2cm}
\noindent\textbf{Results.}
As shown in \cref{table:generation}, MuseTok outperforms both baselines on objective metrics, demonstrating effective harmonic and rhythmic continuation.
On subjective metrics, it lags behind REMI and AMT on Pitch, producing more out-of-key notes that also affect Harmony perception.
% %
% These mistakes come from the imperfect second-stage generation process, which generates music events sequences from codes, where even VAE with continuous latent vectors achieves only around 80\% accuracy.
% %
However, it performs comparably on Structure and Development, matching REMI and surpassing AMT, highlighting its potential for developing musical ideas through code generation.

Beyond these metrics, our two-stage generation model is more robust with long-context primers. %\footnote{Samples available on the demo page.}.
Encoding 16-bar music with REMI+ produces $\sim 800$ events, whereas our code sequence remains fixed at 256 codes when $D$=$16$, supporting long-term generation.
However, this fixed depth can introduce noise for simpler primers.
From \cref{table:tokenization}, 
% As shown in \cref{fig:reconstruction}, 
monophonic pieces reconstruct well with only 8 codes, with extra codes may act as a bias during generation.
These results highlight the need for adaptive quantization and generation strategies for varying musical complexity.

% \begin{figure}[t]
%  \centerline{
%  \includegraphics[width=0.65\columnwidth]{figures/reconstruction_layer.png}}
%  \vspace{-0.2cm}
%  \caption{The reconstruction accuracy of the embedding by aggregating the first $d$ code embeddings across samples in three texture groups.}
%  \label{fig:reconstruction}
%  \vspace{-0.5cm}
% \end{figure}

\subsection{Music Understanding}\label{subsec:understanding}

\textbf{Model Settings and Baselines.}
The melody extraction task is evaluated on POP909~\cite{pop909} using a 3-layer, 4-head Transformer with 128 hidden dim. as classifier, compared with Bi-LSTM (RNN)~\cite{RNN} and MIDI-BERT~\cite{MIDI-BERT} trained on REMI~\cite{remi} sequences.
The chord recognition task is evaluated on POP909 with 133 labels (11 qualities $\times$ 12 roots + ``no chord''), using a 2-layer MLP with 256 hidden dim. as classifier. 
An RNN on REMI+ with bar-level prediction is compared.
The emotion recognition task is conducted on EMOPIA~\cite{emopia} with a 2-layer MLP with 256 hidden dim., compared against RNN, MIDI-BERT~\cite{MIDI-BERT} and MusicBERT~\cite{MusicBERT}. %and a CNN-based piano-roll model~\cite{melodyCNN}.
For this task, MuseTok is retrained with velocity included in the REMI+ encoding.
\\

\vspace{-0.2cm}
\noindent\textbf{Results.}
Classification accuracies are reported in \cref{table:understanding}.
Our model outperforms all baselines on emotion recognition, demonstrating the ability of MuseTok to capture song-level semantic information.
The learned codes also excel at modeling harmony in the challenging 133-class chord recognition task.
Lower performance on melody extraction, along with out-of-key pitch generation above, highlights the need for improved melody modeling of tokenization.

\begin{table}[t]
\small\centering
\resizebox{0.98\linewidth}{!}{
\begin{tabular}{l|cc|cccc}
\toprule
                        & \multicolumn{2}{c|}{\textbf{Objective}} & \multicolumn{4}{c}{\textbf{Subjective}} \\
                        & $\operatorname{sim}_{\mathrm{chr}}$ & $\operatorname{sim}_{\mathrm{grv}}$ & Pit. & Str. & Har. & Dev. \\ 
\midrule
REMI~\cite{remi}        & 94.61                 & \underline{87.41} & \textbf{4.099}    & \textbf{3.927}    & \textbf{3.927}    & \textbf{3.646}    \\
AMT~\cite{Anticipatory} & \underline{94.72}     & 84.08             & \underline{3.839} & 3.328             & 3.516             & 3.156             \\
MuseTok                 & \textbf{95.19}        & \textbf{88.77}    & 3.698             & \underline{3.839} & \underline{3.604} & \underline{3.635} \\ 
\bottomrule
\end{tabular}
}
\vspace{-0.1cm}
\caption{Objective and subjective evaluations on music continuation.}
\label{table:generation}
\end{table}

\begin{table}[t]
\small\centering
\resizebox{0.75\linewidth}{!}{
\begin{tabular}{l|ccc}
\toprule
                            & \textbf{Melody} & \textbf{Chord} & \textbf{Emotion} \\ 
\midrule
% CNN~\cite{melodyCNN}        & -               &  -             & 60.00            \\
RNN~\cite{RNN}              & 89.98           &  38.03             & 53.46            \\
MIDI-BERT~\cite{MIDI-BERT}  & \textbf{90.97}           &  -             & 67.74            \\ 
MusicBERT~\cite{MusicBERT}  & -               &  -             & 77.78            \\
MuseTok                     & 81.92           & \textbf{49.87}         & \textbf{78.95}   \\ 
\bottomrule
\end{tabular}}
\vspace{-0.1cm}
\caption{Classification accuracies on three understanding tasks.}
\label{table:understanding}
\vspace{-0.35cm}
\end{table}

\begin{figure}[t]
 \centerline{
 \includegraphics[width=\columnwidth]{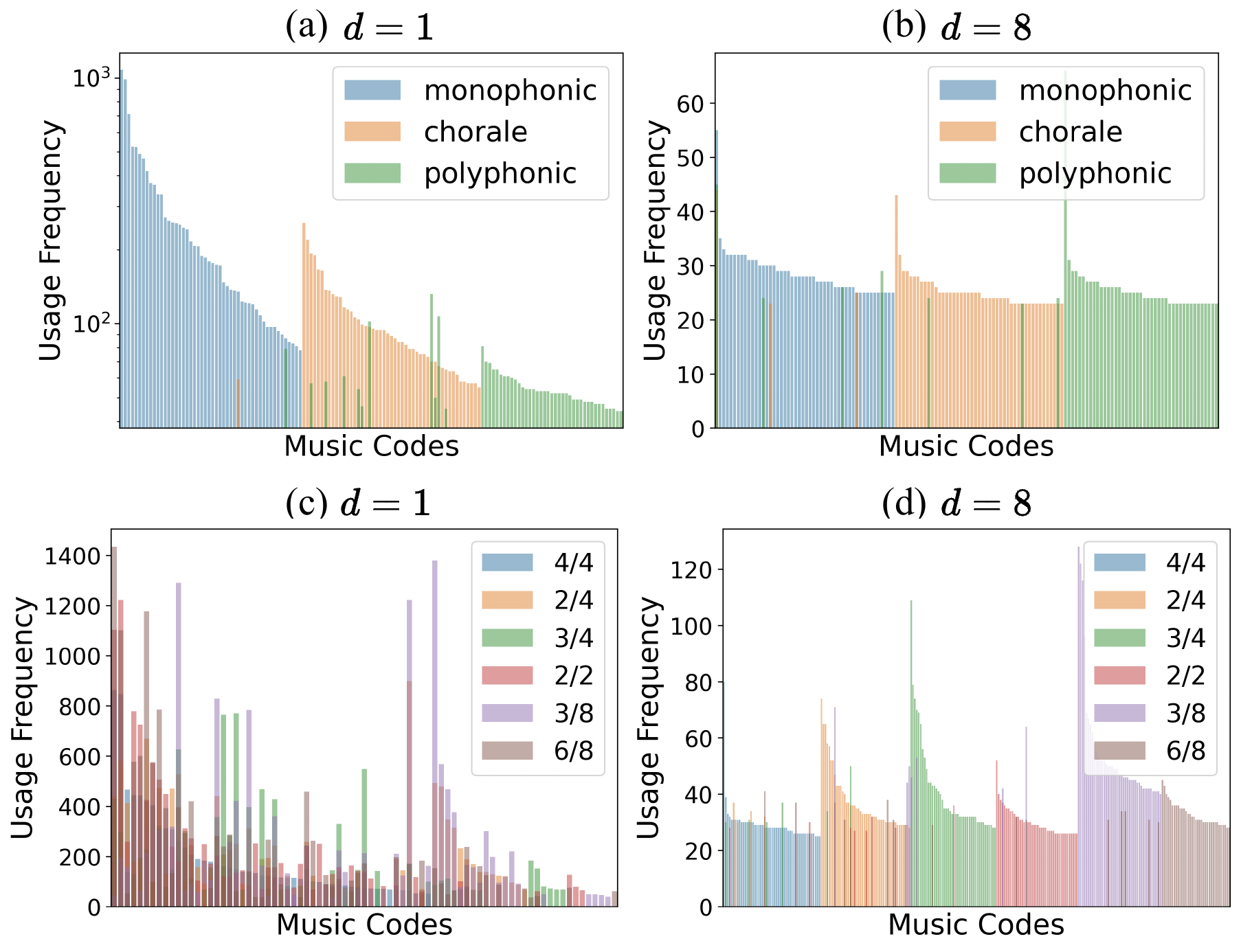}}
 \vspace{-0.2cm}
 \caption{Top-50 used codes across three texture groups, or across six time signatures in the first ($d$=1) and last ($d$=8) codebook.}
 \label{fig:frequency}
\end{figure}
\vspace{-0.2cm}

\begin{figure}[t]
 \centerline{
 \includegraphics[width=0.75\columnwidth]{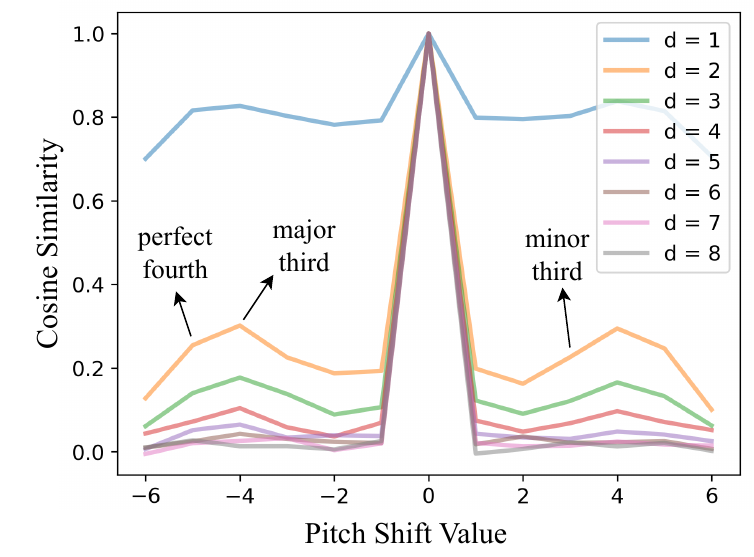}}
 \vspace{-0.2cm}
 \caption{The cosine similarity of code embeddings ($y$-axis) between original and transposed samples by pitch shifts ($x$-axis) across codebooks. Different colored lines denote different codebooks.}
 \label{fig:pitch_shift}
 \vspace{-0.3cm}
\end{figure}

\subsection{How MuseTok Learns Music?}

To explore the musical concepts learned by MuseTok, we conduct two case studies on MuseTok-Small: code usage frequencies across music groups and code embedding similarities on synthetic datasets.
\\

\vspace{-0.2cm}
\noindent\textbf{Code Usage Frequency.}
\cref{fig:frequency} (a) and (b) present the top-50 most frequent codes used from 1000 random samples of monophonic, chorale and polyphonic textures at the first ($d$=1) and last ($d$=8) codebooks.
The frequent code sets are largely distinct across textures, indicating that MuseTok employs different codes and embeddings to represent different textures, treating them as distinct motif and structure development. 
This differentiation persists across all codebooks, as further illustrated on the demo website.

A similar analysis across six time signatures is shown in \cref{fig:frequency} (c) and (d).
Unlike textures, MuseTok almost omits the time signature difference in the first codebook, but gradually diverges in deeper ones, suggesting that different codebooks are allocated to capture different aspects of musical knowledge, with the first one focusing on shared information beyond time signature.
\\

\vspace{-0.2cm}
\noindent\textbf{Embedding Similarity.}
This study examines how code embeddings change across codebooks when applying pitch shifting (key transposition) on music samples.
\cref{fig:pitch_shift} illustrates the cosine similarity of code embeddings ($y$-axis) between original and transposed samples across all 8 codebooks, where transposed samples are generated by shifting all notes from -6 to 6 semitones ($x$-axis).
Embeddings in the first codebook ($d$=1) maintain over 70\% similarity across transpositions, while deeper codebooks gradually diverge. This indicates that (1) transposition-invariant attributes, such as rhythmic information (onset, duration) and relative melodic contour, are mainly processed in earlier codebooks, and (2) absolute pitch information is further processed in deeper codebooks. Although attributes are not fully disentangled, MuseTok demonstrates an unsupervised ability to separate musical concepts across codebooks via data-driven learning.

Another observation relates to musical intervals. Across all codebooks, the highest similarity scores (excluding zero-shift) occur at $\pm 4$ (major third), followed by $\pm 5$ (perfect fourth) and $\pm 3$ (minor third), while $\pm 6$ (augmented fourth) yield the lowest. This suggests that MuseTok captures interval concepts and their relative prevalence in music (major third, perfect forth, and minor third are commonly-used development), as well as interval symmetry, shown by consistent rankings for ascending and descending shifts.

These observations suggest that MuseTok effectively captures fundamental musical concepts, including rhythm, texture, and interval, even without explicit supervision from musical annotations. 
% align closely with human musical perception, as 

% \vspace{-0.25cm}
\section{Conclusion}\label{sec:conclusion}
In this paper, we introduce MuseTok, a discrete representation learning framework for symbolic music. 
RQ-VAE is applied on bar-wise music segments to learn music codes with high-fidelity reconstruction capability.
We investigate the quality of learned codes through symbolic music generation and classification tasks in multiple perspectives, showing its effectiveness on both content generation and semantic understanding.
Further qualitative analyses reveal the underlying musical concepts learned by MuseTok.
In the future, we wish to focus on adaptive tokenization methods, leading to better music generation performance across all music groups.

% \hjy{Dataset statistics, detailed settings and results of subjective listening test, plots with all quantization blocks, long-context generation samples are presented in the demo page}

\vfill\pagebreak
\footnotesize
\bibliographystyle{IEEEbib}
\bibliography{refs}

\end{document}